# Development and Validation of a Sharp Interface Immersed Boundary Method for High-Speed Flows

Punit Pandey[1], Ankit Bansal[1*], Krishna Mohan Singh[1] and Yannick Hoarau[2]

[1] Department of Mechanical and Industrial Engineering

Indian Institute of Technology Roorkee, Roorkee, 247667, India

[2] ICube Laboratory, University of Strasbourg

UMR 7357, CNRS, Strasbourg, France

punit_p@me.iitr.ac.in, ankit.bansal@me.iitr.ac.in[*], krishna.singh@me.iitr.ac.in, hoarau@unistra.fr

**Abstract**

This study presents an advanced sharp-interface immersed boundary method (IBM) integrated with the blastFOAM library on the OpenFOAM platform for high-speed compressible flow simulations. The developed solver extends the existing IBM techniques available in OpenFOAM to compressible regimes, tackling challenges such as shock waves, expansions, and dynamic geometries without needing body-fitted meshes. A novel contribution of this work is the implementation of a slip boundary condition for velocity at immersed surfaces, specifically designed to handle inviscid high-speed flows. The method also combines the second-order polynomial IBM reconstruction with multiple flux schemes such as Kurganov, Tadmor, HLL (Harten–Lax–van Leer), and AUSM+up (Advection Upstream Splitting Method Plus Upwind). The technique achieves significant accuracy across diverse high-speed flow conditions. Extensive validation is performed through supersonic flow cases over a wedge, a cylinder, an aerofoil, a sphere, and a moving piston. Results show excellent agreement with analytical and body-fitted solutions, with sharp resolution of shocks, minimal numerical oscillations, and shock reflections. A grid convergence study confirms the solver's reliability across varying mesh resolutions, while three-dimensional simulations highlight its capability for scaled-up applications. This solver provides a flexible, efficient, and accurate tool for capturing high-speed flow phenomena across various Mach numbers and geometries. It offers significant advantages in mesh handling, particularly for dynamic or intricate configurations, making it ideal for aerospace and engineering applications involving compressible flows.

**Nomenclature**

| | | | |
|---|---|---|---|
| **U** | Vector of conservative variables | GCIBM | ghost-cell-based immersed boundary method |
| **F** | Matrix of convective fluxes | | |
| **F**$_f$ | Intercell flux | IB | Immersed Boundary |
| $p$ | Pressure | Ma | Mach number |
| $L$ and $R$ | Left and right states | | |
| h | Mesh size | **Greek Letters** | |
| $M$ | Mach number | $\rho$ | Density |
| $V_i$ | Arbitrary control volume | $\gamma$ | Specific heat ratio |
| $w_i$ | Weight for cell $i$ | $\phi$ | Primitive variable |
| $r_i$ | Distance of cell $i$ from IB cell | $\Omega_i$ | Surface of the control volume $i$ |
| | | $\delta$ | Angle between the vector from the IB point to cell $i$ and the local IB normal. |
| **Abbreviations** | | | |
| IBM | Immersed Boundary Method | $\eta$ | Angle factor |
| HLL | Harten–Lax–van Leer | $\theta$ | Shock angle |
| AUSM+up | Advection Upstream Splitting Method Plus Upwind | $\zeta$ | Generic flow variable used for error evaluation |
| CFD | Coputational Fluid Dynamics | $\emptyset$ | Volumetrix flow rate through a cell face |
| STL | Stereolithography | $\|\varepsilon\|_2$ | L$_2$-norm of the discretization error |

## 1. Introduction

Numerical modeling of high-speed flows is challenging as it requires managing shock waves, pronounced temperature and density gradients, and notable numerical instabilities. It becomes even more difficult when such simulations must deal with complex and moving geometries. These hurdles are conventionally addressed using body-fitted grids with mesh regeneration, which can be computationally intensive. Effective alternatives to mitigate these issues are the meshless methods and the immersed boundary methods (IBM). The IBMs employ a non-body conforming grid strategy. The concept of the IBM was initially introduced by (Peskin 1977) for simulating blood flow dynamics within the heart. Since its inception, IBM has seen extensive development efforts to enhance accuracy and computational efficiency. Cui et al. (2017) categorize IBMs into two principal types based on the representation of fluid-structure interfaces: diffused-interface methods and sharp-interface methods. In the diffused interface IBMs, the interface is treated using a continuous forcing approach, where the momentum equation is modified by adding a forcing term. This results in the smearing of the actual boundary due to the distribution of forcing terms (Constant et al. 2017; Huang and Tian 2019) and source terms (Riahi et al. 2018) to neighboring Cartesian grid points (Yang et al. 2009). Numerous variants of diffused-interface IBMs have been devised for applications ranging from simulating compressible flow over moving bodies to handling compressible multiphase flows (Wang et al. 2017;

Wang et al. 2020). One notable advantage of diffuse-interface IBMs lies in their formulation, which remains independent of spatial discretization methods and facilitates straightforward integration into existing fluid solvers. Despite these strengths, (Sotiropoulos and Yang 2013) highlighted the challenges associated with classic diffuse-interface IBMs when applied to fluid-structure interaction problems. The stiffness inherent in the forcing terms can lead to numerical instabilities and unwanted oscillations, posing significant obstacles in simulations of such complex interactions.

An alternative method that avoids the direct addition of discrete forces into the governing equations was introduced by Mohd-Yusof (1997), and Fadlun et al. (2000). In this method, the no-slip boundary condition is enforced at the interior fluid cells near the immersed boundary. The values of the flow variables at these fluid cells are obtained through interpolation along the grid lines between the closest interior cells. Further, the momentum equation is modified in these fluid cells to implement the physical boundary conditions. This technique is commonly classified as a Cartesian immersed boundary method.

In contrast to the diffused-interface method, the sharp-interface IBMs are characterized by a sharp boundary representation. The sharp interface method is also called the discrete forcing approach. This approach encompasses several methodologies, such as the cut-cell method (Schneiders et al. 2016), the Cartesian IB method (Gilmanov and Sotiropoulos 2005; Cui et al. 2017; Brahmachary et al. 2020), and the ghost-cell method (Qu et al. 2017; Khalili et al. 2018). Ghias et al. (2006) developed one of the first sharp-interface immersed boundary methods for computing viscous, subsonic compressible flows with complex stationary immersed boundaries. Later, De Tullio et al. (2007) introduced an IBM for compressible flows to simulate complex flow fields over a broad range of Mach numbers. They incorporated local grid refinement to enable higher resolution of flow features. Hartmann et al. (2010) implemented a Cartesian cut-cell method capable of solving two and three-dimensional viscous, compressible flow problems on arbitrarily refined, graded meshes, offering flexibility in handling diverse flow regimes. Similarly, Schneiders et al. (2012) developed a cut-cell method for Cartesian meshes for simulating viscous compressible flows with moving boundaries, enhancing the ability to handle dynamic geometries. Nam and Lien (2014) proposed a ghost-cell-based immersed boundary method (GCIBM) for simulating compressible turbulent flows around complex geometries. This approach enforces boundary conditions using ghost cells located inside the immersed solid. Additionally, Brehm et al. (2015) presented a higher-order finite difference IBM for solving compressible Navier-Stokes equations. They applied the no-slip boundary conditions to the immersed boundary for improved accuracy in capturing the boundary layer effects.

Among all sharp-interface IBMs, the cut-cell method stands out for its ability to clearly delineate the interface. However, the cut-cell method faces challenges due to the intricate cell reshaping procedures, particularly in simulating complex problems involving moving bodies. In contrast, the Cartesian IBM treats the fluid points close to the immersed body as IB points, whereas the ghost-cell IBM represents the boundary by treating solid mesh points adjacent to the immersed boundary as ghost points. Tran and Plourde (2014) employed both methods effectively in internal flow simulations, highlighting their comparative ease in point recognition procedures and flux calculations around immersed boundaries. Khalili et al. (2018) extended the ghost-cell IBM to model viscous compressible flows, employing bilinear interpolation techniques for variable reconstruction. For simulating high-speed compressible inviscid flows, a similar ghost-cell IBM has been used by Wang et al. (2020). The inherent challenge of the ghost-cell method lies in its accuracy in interpolating flow variables. In contrast, the reconstruction procedure for flow variables in the Cartesian IBM is simple, straightforward, and more accurate, as it avoids the use of data inside the solid domain.

Most immersed boundary methods have focused on incompressible flows (Choi et al. 2006) in biomedical applications, microfluidics, etc. The applicability of the IBM to compressible flows, particularly for high-speed flows, is still in the developing phase. The majority of studies have utilized the ghost-cell methodology (Brahmachary et al. 2017). To model subsonic and transonic flows, Zhang and Zhou (2014) adopted the Cartesian immersed boundary method of Gilmanov and Sotiropoulos (2005). In the IBM of Gilmanov and Sotiropoulos (2005), the flow variables in the IB cells are interpolated along the perpendicular line between the immersed boundary point and the nearest fluid cell center. Tukovic and Jasak (2012) employed quadratic interpolation and an inverse distance weighted least squares method to reconstruct flow variables at the IB cells for incompressible flow. The details of the least squares procedure can be found in Seo and Mittal (2010). In our previous study (Pandey et al. 2025), we extended the discrete forcing IB framework of Tukovic and Jasak (2012) to simulate high-speed compressible flows over stationary bodies. We employed the blastFOAM solver to solve the compressible Navier-Stokes equations discretized on a Cartesian grid.

The blastFOAM is an open-source C++ library (available in OpenFOAM) that simulates multiphase high-speed reacting flows. The blastFOAM implements 2nd and 3rd-order time marching schemes, equation of states, and flux schemes such as HLL, Kurganov, Tadmor, and AUSM+up (Heylmun et al. 2021). In the current study, we have developed and applied the coupled blastFOAM-IBM solver to comprehensively understand the effect of different flux discretization schemes in different Mach number regimes for various stationary and moving boundary problems. The selection of sharp-interface IBM is important because it provides precise prediction of shock wave locations, which is critical in our investigation.

## 2. Methodology

This section outlines the governing equations and the simulation methodology employed for high-speed inviscid compressible flows. The compressible flow solver used in this study is based on the blastFOAM library within the open-source CFD framework OpenFOAM. blastFOAM offers a high-fidelity algorithm suite for solving the single-phase and multi-phase compressible Navier–Stokes equations. It supports a wide range of numerical schemes, including advanced flux discretization techniques (such as Riemann solvers), high-order temporal integration methods like Runge–Kutta schemes, and multiple equations of state. Additionally, it incorporates slope-limiter and higher-order interpolation schemes for accurate variable reconstruction. In this study, the IBM formulation of Tuković and Jasak (2012) is extended for high-speed flows by integrating and enhancing the functionalities of blastFOAM and foam-extend-4.0. A detailed description of the numerical methods and implementation strategies is provided in the subsequent sections.

### 2.1. Governing equations

The Euler equations govern the inviscid compressible flow of an ideal gas. These equations are expressed in conservative form as

$$\frac{\partial \mathbf{U}}{\partial t} + \nabla \cdot \mathbf{F} = 0 \tag{1}$$

Here, **U** is the vector of conservative variables and **F** is the matrix of convective fluxes, which is given by

$$\mathbf{U} = \begin{bmatrix} \rho \\ \rho \mathbf{u} \\ \rho E \end{bmatrix} \quad \mathbf{F} = \begin{bmatrix} \rho \mathbf{u} \\ \rho \mathbf{u} \otimes \mathbf{u} + p\mathbf{I} \\ (\rho E + p)\mathbf{u} \end{bmatrix} \tag{2}$$

In Eq. (2), $p$, $\rho$, **u**, **I**, and $\otimes$ are the pressure, density, velocity vector, identity matrix, and the tensor product operator, respectively. Further, $\rho E$ is the total energy per unit volume given by

$$\rho E = \frac{1}{2}\rho \|\mathbf{u}\|^2 + \rho e \tag{3}$$

where $e = p/((\gamma - 1)\rho)$ is the specific internal energy and $\gamma$ is the specific heat ratio.

Equation (1) is integrated over an arbitrary control volume $V_i$ bounded by surface $\Omega_i$ as:

$$\frac{\partial}{\partial t}\int_{V_i} \mathbf{U}\, dV + \int_{V_i} \nabla \cdot \big(\mathbf{F}(\mathbf{U})\big)\, dV = 0 \tag{4}$$

Using the Gauss divergence theorem, we get

$$\frac{\partial}{\partial t}\int_{V_i} \mathbf{U}\, dV + \int_{\Omega_i} \big(\mathbf{F}(\mathbf{U})\big) \cdot d\mathbf{A} = 0 \tag{5}$$

In a cell-centred finite-volume dicretisation, we define the cell average

$$\overline{\mathbf{U}}_i(t) = \frac{1}{V_i} \int_{V_i} \mathbf{U}(\mathbf{x}, t) dV \qquad (6)$$

where $V_i$ is the volume of cell $i$. Approximating the surface integral by a sum of face contributions using the midpoint (face-centered) quadrature gives

$$\int_{\Omega_i} \bigl(\mathbf{F}(\mathbf{U})\bigr) \cdot d\mathbf{A} = \int_{\Omega_i} \bigl(\mathbf{F}(\mathbf{U})\bigr) \cdot \mathbf{n}\, dA = \sum_{f \in \partial \Omega_i} \mathbf{F}_f A_f \qquad (7)$$

where $\mathbf{F}_f$ is the numerical flux evaluated at face $f$ and $A_f$ is the face area. Using an explicit first-order time discretisation (forward Euler) for the time derivative

$$\frac{\partial}{\partial t}\overline{\mathbf{U}}_i = \frac{\overline{\mathbf{U}}_i^{n+1} - \overline{\mathbf{U}}_i^n}{\Delta t} \qquad (8)$$

we obtain the finite-volume update

$$(\overline{\mathbf{U}})_i^{n+1} = (\overline{\mathbf{U}})_i^n - \frac{\Delta t}{V_i} \sum_{f \in \Omega_i} \mathbf{F}_f A_f \qquad (9)$$

We employ the approximate Riemann solvers (HLL and AUSM+up) and central schemes (Kruganov/Tadmor) for calculating conservative fluxes at cell interfaces ($\mathbf{F}_f$) (Toro et al. 1994). Based on the interpolated left and right-hand states ($\mathbf{U}_L$ and $\mathbf{U}_R$), the Riemann problem can be approximated in the region between the waves ($S_L$ and $S_R$), as shown in Fig. 1. The $L$ and $R$ states are obtained using a combination of base interpolation schemes (linear or cubic) and various slope limiters such as upwind, minmod, vanLeer, superbee, etc. The flux schemes are described in detail in the Appendix A.

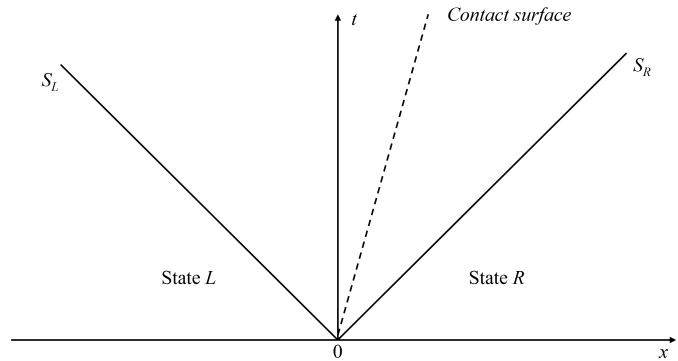

**Fig. 1.** Riemann problem representation between the states $L$ and $R$

## 2.2. Sharp interface immersed boundary method

This section presents the formulation and implementation of the sharp-interface Immersed Boundary Method (IBM). The sharp-interface IBM emphasizes maintaining a clear and geometrically accurate representation of the solid-fluid interface, ensuring this interface remains sharply defined throughout the numerical computations. Here, the computational domain is discretized using a background

Cartesian mesh, where solid bodies are immersed without requiring the mesh to conform to the geometry. One of the central challenges in IBM is the accurate imposition of boundary conditions. The immersed geometry is represented by an unstructured triangulated surface mesh. This mesh consists of a collection of triangular elements that define the surface of the solid body embedded in the computational domain. A standard method for representing such a triangulated surface is through an element-to-vertex connectivity data structure, where each triangular element is described by three vertices. These vertices specify the spatial coordinates of the triangle's corners, thereby defining its shape and position in three-dimensional space. This type of data structure is commonly used in file formats such as STL (stereolithography), where each triangle is listed along with its corresponding vertex coordinates and an optional surface normal vector.

The implementation follows a two-step process involving the identification of relevant mesh cells and the reconstruction of flow variables to accurately enforce boundary conditions. The detailed methodology for each step is presented in the subsequent sections.

**2.2.1. Identification**

This step classifies the cells into solid, fluid, and immersed boundary (IB) cells. This step plays a crucial role in ensuring the accuracy and efficiency of the simulation. IB cells are those fluid cells near the fluid-solid interface that are intersected by the immersed solid boundary. This classification is critical because it ensures that fluid cells do not share faces with solid cells, a requirement that becomes especially important in moving body simulations.

Further, we need to identify an extended stencil of fluid cells for interpolating the field variables and implementing boundary conditions at IB cells. For the IB cell P, the corresponding extended stencil $(S_i)^P$ can be determined as shown in Fig. 2. The selection of extended stencil has been explained by Tukovic and Jasak (2012). There are three key parameters for selecting cells of the extended stencil, namely angle factor ($\eta$), radius factor, and maximum cell rows. The angle factor parameter defines a cosine-based rejection threshold such that cells are accepted if the angle between the inward surface normal and the cell direction vector satisfies $\delta \leq 180° - \eta$.

The radiusFactor determines the size of the search region around the immersed boundary cells. The larger the radiusFactor, the more cells will be in the extended stencil. Lastly, the third parameter defines the maximum number of rows of neighboring cells included in the extended stencil. This parameter limits the depth of the search for neighboring cells. Extending the search to second or third-layer neighboring cells can enhance interpolation accuracy but raises computational costs as more cells are involved in the calculations.

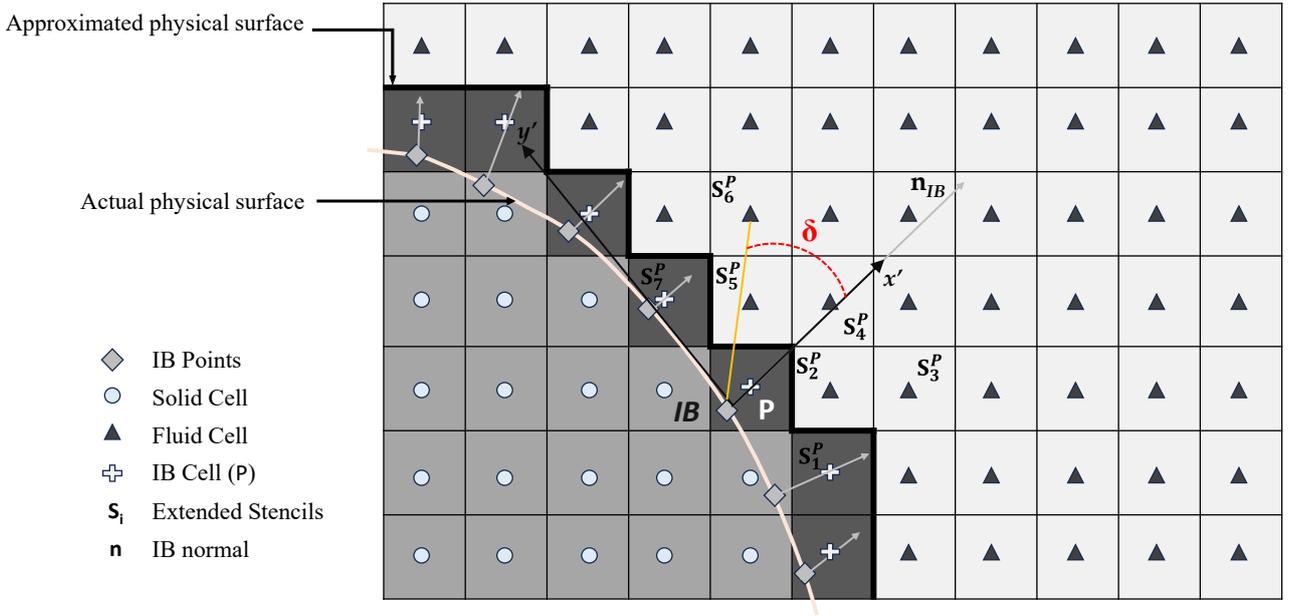

**Fig. 2.** Schematic for ndentification of cells and selection of extended stencil

### 2.2.2. Reconstruction of variables at IB cells

In the fluid region, the flow field is simulated using the standard finite volume discretization, while the solid cells are excluded from the computation. The flow variables in the IB cells are reconstructed through an interpolation scheme using the extended stencil of fluid cells. The reconstruction of variables at an IB cell begins by determining the normal vector to the immersed boundary passing through the center of the IB cell represented as $\mathbf{n}_{IB}$ in Fig. 2. The intersection of the normal vector with the immersed boundary is identified as the IB point. The reconstruction procedure employs the extended stencils (marked with points $(S_i)^P$). The specified boundary conditions are strictly imposed at the IB points. This ensures that boundary conditions are accurately transferred to the computational domain, thereby preserving the integrity of the flow solution near the immersed boundary.

In this work, we have employed the following quadratic interpolation to implement the Dirichlet boundary condition

$$\phi_P = \phi_{IB} + C_0\bar{x} + C_1\bar{y} + +C_2\bar{z} + C_3\bar{x}\bar{y} + C_4\bar{y}\bar{z} + C_5\bar{z}\bar{x} + C_6\bar{x}^2 + C_7\bar{y}^2 + C_8\bar{z}^2 \qquad (10)$$

where $\bar{x} = (x_P - x_{IB})$, $\bar{y} = (y_P - y_{IB})$, $\bar{z} = (z_P - z_{IB})$, and $\phi$ is any primitive variable. The subscript *P* and *IB* signify the immersed cell's center and the IB point, respectively. The variables *x*, *y*, and *z* are in the global coordinate system.

Similarly, the reconstruction of the Neumann boundary conditions follows a similar procedure given in Eq. (11). Here a local coordinate system ($x'\ y'\ z'$) is defined for in way, $x'$ aligns with the local normal ($\mathbf{n}_{IB}$) (Singh et al., 2015).

$$\begin{aligned}\phi_P = D_0 + [\mathbf{n}_{IB} \cdot (\nabla\phi)_{IB}]x'_P + D_1 y'_P + D_2 z'_P + D_3 x'_P y'_P + D_4 y'_P z'_P + D_5 z'_P x'_P \\ + D_6 (x'_P)^2 + D_7 (y'_P)^2 + D_8 (z'_P)^2\end{aligned} \qquad (11)$$

where $\mathbf{n}_{IB}$ is the normal vector at the IB point.

For compressible inviscid flows past the solid bodies the we have applied the Neumann boundary condtion for temperature and pressure and slip condition for velocity at the immersed surface. The implementation of the Neumann condition can be done as explained in Eq. (11), but the slip condition, which involves the mix of Dirichlet $u_{x'} = 0$ to ensure no penetration at the surface and Neumann boundary conditions, $\frac{\partial u_{y'}}{\partial x'} = 0$ and $\frac{\partial u_{z'}}{\partial x'} = 0$ on the immersed surface which can be considered as a Robin-type boundary condition. For the slip, we need to decompose the velocity vector into normal and tangential components with respect to the local IB normal. Here, the $y'$ and $z'$ are the two tangential directions at the IB point as shown schematically in Fig. 3. These velocity components are given by

$$u_{x'_P} = u_{x'_{IB}} + C_0\bar{x} + C_1\bar{y} + +C_2\bar{z} + C_3\bar{x}\bar{y} + C_4\bar{y}\bar{z} + C_5\bar{z}\bar{x} + C_6\bar{x}^2 + C_7\bar{y}^2 + C_8\bar{z}^2 \qquad (12)$$

$$\begin{aligned}u_{y'_P} &= D_0 + [\mathbf{n}_{IB} \cdot (\nabla u_{y'})_{IB}]x'_P + D_1 y'_P + D_2 z'_P + D_3 x'_P y'_P + D_4 y'_P z'_P + D_5 z'_P x'_P \\ &\quad + D_6(x'_P)^2 + D_7(y'_P)^2 + D_8(z'_P)^2\end{aligned} \qquad (13)$$

$$\begin{aligned}u_{z'_P} &= D_0 + [\mathbf{n}_{IB} \cdot (\nabla u_{z'})_{IB}]x'_P + D_1 y'_P + D_2 z'_P + D_3 x'_P y'_P + D_4 y'_P z'_P + D_5 z'_P x'_P \\ &\quad + D_6(x'_P)^2 + D_7(y'_P)^2 + D_8(z'_P)^2\end{aligned} \qquad (14)$$

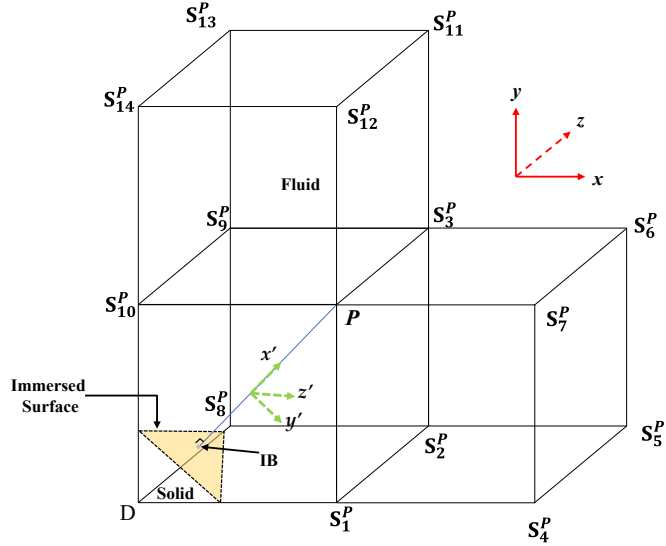

**Fig. 3.** Schematic of velocity component decomposition

The coefficients $C_\beta$ and $D_\beta$ are determined using a weighted least squares method (Tuković and Jasak, 2012) which involves solving a system of linear equations

$$\mathbf{A}\mathbf{x} = \mathbf{b} \qquad (15)$$

Here, $\mathbf{b} \in \mathbb{R}^n$ is the vector of known field values at neighbouring points points, $\mathbf{x} \in \mathbb{R}^m$ is the coefficient vector (unknowns to be solved) and for each IB cell the matrix $\mathbf{A} \in \mathbb{R}^{n \times m}$ is formulated in which the rows represent the number of neighbouring cells ($n$) whereas the columns define the polynomial basis ($m$).

For the 2-D Dirichlet immersed boundary reconstruction, the quadratic interpolation polynomial contains five unknown coefficients. Therefore, at least five neighbouring fluid points are required for the weighted least-squares system. For an $n$-point stencil, the matrix **A** is defined as

$$\mathbf{A} = \begin{bmatrix} X_1 & Y_1 & X_1Y_1 & X_1^2 & Y_1^2 \\ X_2 & Y_2 & X_2Y_2 & X_2^2 & Y_2^2 \\ \vdots & \vdots & \vdots & \vdots & \vdots \\ X_n & Y_n & X_nY_n & X_n^2 & Y_n^2 \end{bmatrix} \tag{16}$$

Similarly, for 3-D, at least 9 neighbouring points must be identified. Hence it can expressed as

$$\mathbf{A} = \begin{bmatrix} X_1 & Y_1 & Z_1 & X_1Y_1 & Y_1Z_1 & Z_1X_1 & X_1^2 & Y_1^2 & Z_1^2 \\ X_2 & Y_2 & Z_2 & X_2Y_2 & Y_2Z_2 & Z_2X_2 & X_2^2 & Y_2^2 & Z_2^2 \\ \vdots & \vdots & \vdots & \vdots & \vdots & \vdots & \vdots & \vdots & \vdots \\ X_n & Y_n & Z_n & X_nY_n & Y_nZ_n & Z_nX_n & X_n^2 & Y_n^2 & Z_n^2 \end{bmatrix} \tag{17}$$

Where $X_i = x_{IB} - x_i$, $Y_i = y_{IB} - y_i$, and $Z_i = z_{IB} - z_i$, with $i$ being the $i^{th}$ neighbouring cell. In the weighted least-square error method, a weighting function modifies the least squares system to ensure better accuracy. Thus the minimization problem is written as

$$\min_x \|\mathbf{w}(\mathbf{Ax} - \mathbf{b})\|^2 \tag{18}$$

which simiplifies to

$$\mathbf{A^T w A x} = \mathbf{A^T w b} \tag{19}$$

where $\mathbf{w} \in \mathbb{R}^{n \times n}$ is a diagonal matrix that depends on the distance of the $i^{th}$ cell centre from the IB cell centre. We have employed the following weight functions for the Dirichlet and Neumann boundary conditions, respectively.

$$w_i = \left[1 - \left(\frac{r_i}{1.1 \times (r_{max})}\right)\right] \text{ and } w_i = \frac{1}{2}\left[1 + \cos\left(\pi \frac{r_i}{r_{max}}\right)\right] \tag{20}$$

where, $r_i$ is the distance of cell $i$ from the coreesponding IB cell and the $r_{max}$ is the distance of the farthest cell in the extended stencil.

### 2.2.3. Discrete mass conservation

To verify the discrete conservation behavior of the current IBM reconstruction, we simulate supersonic flow over a wedge with a semi-vertex angle of 20° at a freestream Mach number of 2. Simulations are carried out on four progressively refined Cartesian grids over a domain of size 0.1×0.15, using unit aspect ratio cells as shown in Table 1.

Mass conservation is assessed by computing the mass loss, defined as

$$Mass_{loss} = \sum \rho \emptyset \quad \text{where} \quad \emptyset = \mathbf{u}_f \cdot \mathbf{S}_f \tag{21}$$

Here $\rho \emptyset$ is the mass flux across the faces in the domain, $\mathbf{u}_f$ is the velocity at a cell interface and $\mathbf{S}_f$ is the face area. On a body-fitted mesh, the mass flux through the solid body surface would be zero, and the discrepancy between inlet and outlet fluxes would primarily reflect the error in mass conservation.

As in this immersed boundary reconstruction, variables in the IB cells are interpolated from nearby cells rather than directly solved, this leads to a finite mass loss even steady state is reached as shown in Fig. 4, also it demonstrates that mass loss decrease with mesh refinement.

**Table 1.** Mesh configurations along with characterstic grid spacings.

| Mesh | h |
|---|---|
| 100 × 150 | 1/100 |
| 150 × 225 | 1/150 |
| 200 × 300 | 1/200 |
| 250 × 375 | 1/250 |

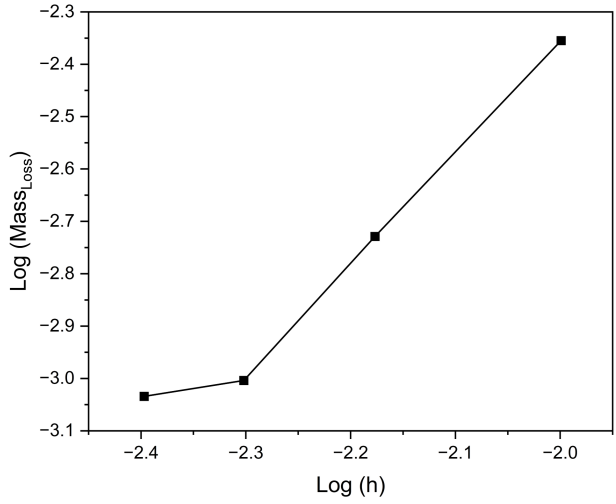

**Fig. 4.** Effect of mesh refinement on mass loss

### 2.2.4. Discrete mass conservation

To assess the order of convergence, we examine the blastIBFOAM solver using a supersonic vortex test case that possesses an analytical solution. This case represents inviscid, isentropic flow confined between two concentric circular arcs.

For numerical simulation, the channel geometry is immersed within a square computational domain of size 1.5×1.5, which is discretized using multiple uniform Cartesian meshes. The channel boundaries correspond to circular arcs of radii 1 and 1.384, as illustrated in Fig. 5(a). The region enclosed by these arcs constitutes the fluid domain, while the space outside is treated as solid. The IB cells form a thin layer separating the fluid and solid regions. The analytical expression for the flow field in this configuration is given as follows:

$$\rho = \rho_i \left[ 1 + \frac{\gamma - 1}{2} \left( 1 - \frac{R_i^2}{r^2} \right) M_i^2 \right]^{\frac{1}{\gamma - 1}} \quad (22)$$

Here, $\rho_i$ and $M_i$ represent the density and Mach number, respectively, at the inner radius $R_i$. This formulation provides the density distribution ($\rho$) as a function of the radial position ($r$). For the current test case, the Mach number at the inner radius is specified as ($M_i = 2.25$), while the corresponding density is set to ($\rho_i = 1$). The velocity components and pressure for this flow field are expressed as follows

$$u = U_{Ri}\frac{yR_i}{r^2}; v = -U_{Ri}\frac{xR_i}{r^2} \tag{23}$$

$$p = p_i\left[1+\frac{\gamma-1}{2}\left(1-\frac{R_i^2}{r^2}\right)M_i^2\right]^{\frac{\gamma}{\gamma-1}} \tag{24}$$

where, $U_{Ri} = M_i a_i$ with $a_i = \sqrt{\gamma p_i/\rho_i}$

we have employed $L_2$ – norm of error for a variable $\zeta$ defined as

$$\|\varepsilon\|_2 = \sqrt{\frac{\sum_{j=1}^{n}(\zeta_j - \zeta_{\text{exact}})^2}{n}} \tag{25}$$

The $L_2$-norm errors in density and pressure were computed across successive mesh refinements, and the corresponding log–log plots yield convergence slopes of approximately 2.2 and 2.25, respectively as shown in Fig. 5(b). These results confirm that the present solver achieves near-second-order spatial accuracy for smooth, continuous flows. Although the quadratic IB reconstruction used in this study is formally third-order accurate, the global accuracy is constrained by the spatial discretization schemes available in blastFOAM.

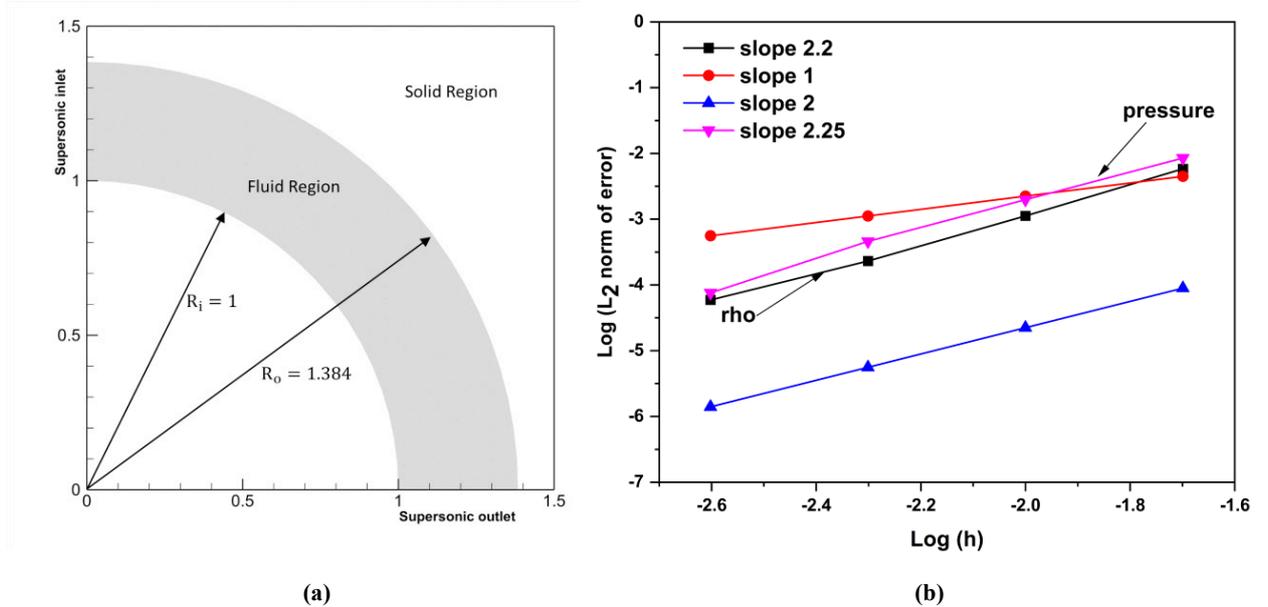

**Fig. 5.** (a) Computational domain and (b) $L_2$ Norm of errors in pressure and density

## 3. Results and discussion

We have simulated various cases of external supersonic flow to validate the newly developed blastIBFOAM solver. These cases are widely employed to validate flux schemes and physical models for compressible flow. We start our validation with a rather simple case of flow over a compression wedge and then proceed toward more challenging cases of two and three-dimensional bluff bodies and moving boundary problems.

### 3.1. Supersonic flow over a wedge

In this case, we compare flux schemes based on the supersonic flow over a 15º wedge for Ma 3.0 and 5.0. The simulation uses a two-dimensional rectangular computational domain of dimensions (0.5 × 0.4). The working fluid is an idealized, inviscid gas characterized by a molar mass of 11,640.3, with initial conditions of $p=1$ and $T=1$. These properties yield a speed of sound of unity. Supersonic inflow conditions are applied at the domain inlet, while a zero-gradient condition is imposed at the outlet to allow unimpeded flow exit. To evaluate the order of convergence of the current immersed boundary solver, simulations were performed for Ma 3.0 flow using four different uniform grids: 200×160, 400×320, 800×640, and 1600×1280. The corresponding sizes are $\Delta h =$ 0.0025, 0.00125, 0.000625, and 0.0003125, respectively. The error in shock angle and pressure was computed with respect to the exact solution as presented in Table 2, and plotted against the mesh size in Fig. 6. It is observed, the order of convergence is approximately 2, which confirms the second-order spatial accuracy of the present immersed boundary solver.

**Table 2.** Shock angle and pressure error

| h | $|\theta_{cal} - \theta_{exact}|$ | $|p_{cal} - p_{exact}|$ |
|---|---|---|
| 0.0025 | 1.849 | 0.302 |
| 0.00125 | 0.836 | 0.0755 |
| 0.00125 | 0.4 | 0.0188 |
| 0.0003125 | 0.03 | 0.0068 |

In supersonic flow conditions, a wedge-shaped geometry generates an attached oblique shock wave at its leading edge. This classical problem has a well-established analytical solution, allowing for direct comparison and validation of numerical solvers. Simulations utilizing a body-fitted mesh were carried out using the numerical flux scheme developed by Kurganov and Tadmor (2000), implemented with in rhoCentralFOAM solver (Greenshields et al., 2009). Temporal integration was handled using a first-order forward Euler scheme. From Fig. 7(a and b), it is observed that the Mach contours obtained from the current IBM solver reveal a sharply defined oblique shock attached precisely at the leading edge of the wedge for both Ma 3.0 and Ma 5.0 cases.

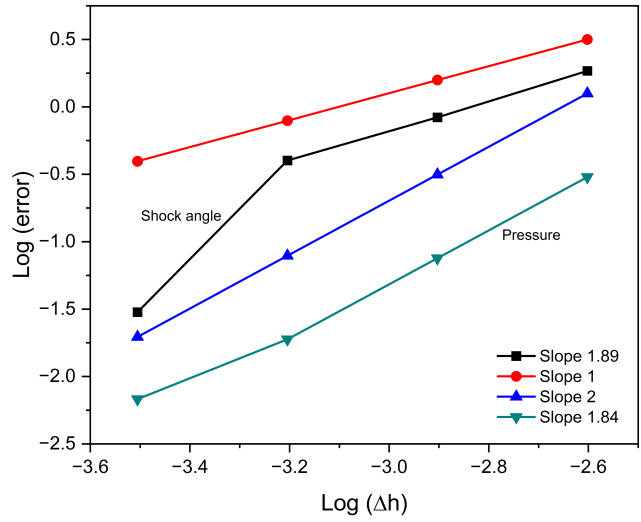

**Fig. 6.** Convergence of shock angle and pressure error versus mesh size

Next, we plotted the density profiles parallel to the base of the wedge, and the obtained results are compared with the body-fitted simulation for both the Mach numbers, as shown in Fig. 7(c and d), respectively. The simulation results show that the onset of shock occurs sharply at the leading edge for both the IBM solver and the body-fitted solver.

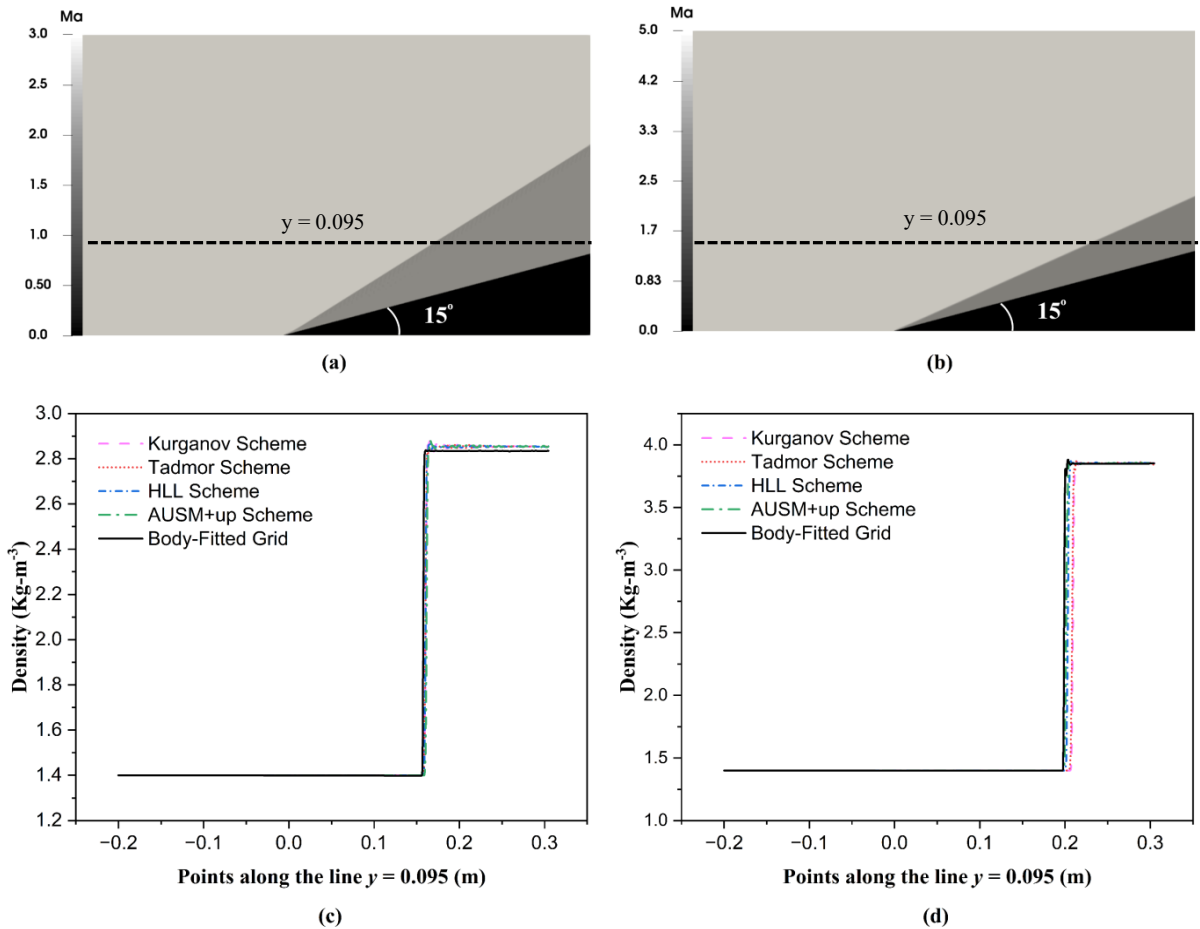

**Fig. 7.** Mach contours for (a) Ma 3.0 and (b) Ma 5.0; Density profiles for (c) Ma 3.0 and (d) Ma 5.0

The current IBM solver can capture the supersonic flow around a compression corner accurately and sharply. The results reveal that all the flux schemes can accurately predict the post-shock values. The Kurganov and Tadmor schemes tend to overpredict the shock location, whereas AUSM+up and HLL schemes seem closer to the body-fitted result; this differentiation can be made in the density profile of Ma 5.0. This comparison highlights the excellent performance of the current IBM solver in capturing the post-shock characteristics and resolving the shock.

Table 3 compares numerical results for the pre-shock and post-shock states from the best flux scheme of the IBM solver with theoretical results. The key parameters evaluated in this study include the post-shock Mach number ($Ma_2$), pressure ratio ($p_2/p_1$), temperature ratio ($T_2/T_1$), density ratio ($\rho_2/\rho_1$), and the shock angle, which is the angle formed between the oblique shock and the horizontal plane. The results produced by the present solver exhibit excellent agreement with the theoretical predictions over the wedge.

**Table 3.** Comparison of current numerical results with the exact solution using supersonic oblique shock relations

| Case | Flow variables | Present | Exact |
|---|---|---|---|
| **$Ma_1=3.0$** | $Ma_2$ | 2.253 | 2.255 |
| | $p_2/p_1$ | 2.818 | 2.822 |
| | $T_2/T_1$ | 1.388 | 1.388 |
| | $\rho_2/\rho_1$ | 2.033 | 2.0324 |
| | Shock angle (Degrees) | 32.2 | 32.24 |
| **$Ma_1=5.0$** | $Ma_2$ | 3.504 | 3.504 |
| | $p_2/p_1$ | 4.783 | 4.7808 |
| | $T_2/T_1$ | 1.7343 | 1.7362 |
| | $\rho_2/\rho_1$ | 2.754 | 2.753 |
| | Shock angle (Degrees) | 24.29 | 24.32 |

The immersed boundary (IB) method introduces additional interpolation operations near the immersed surface, and the associated computational cost depends primarily on the extended stencil depth. To quantify this effect, Ma 3 wedge simulations were performed using 3, 5, and 7 neighboring cell layers under identical numerical settings. The total wall-clock times were 940.65 s, 1020.45 s, and 1154.67 s, corresponding to increases of 8.5% and 22.7% relative to the 3-layer case.

A comparison with a body-fitted solver for the same test case yielded runtimes of 4917.7 s (IB) and 5671.7 s (body-fitted), indicating an overall 13% reduction in computational time for the IB method. The efficiency gain arises from the use of Cartesian meshes, avoidance of highly skewed cells, and elimination of mesh deformation. Therefore, although deeper stencils moderately increase local reconstruction cost, the proposed IB formulation remains computationally advantageous while preserving solution accuracy, demonstrating a favorable balance between accuracy and efficiency.

**3.2. Piston moving with supersonic velocity**

To evaluate the performance and accuracy of the IBM solver for moving bodies, we consider a problem involving a piston moving at a constant Mach number of 2.0 defined with respect to the static initial conditions of the fluid. The schematic of the problem is illustrated in Fig. 8. The piston has a thickness of $L$ and is initially positioned at the center of the two-dimensional domain of $128L \times 4L$. The previous investigations of this problem (Murman et al., 2003; Muralidharan and Menon, 2018) have emphasized that maintaining the strict conservation of mass, momentum, and energy is critical to accurately determining the shock wave speed and capturing the correct characteristics of both regions.

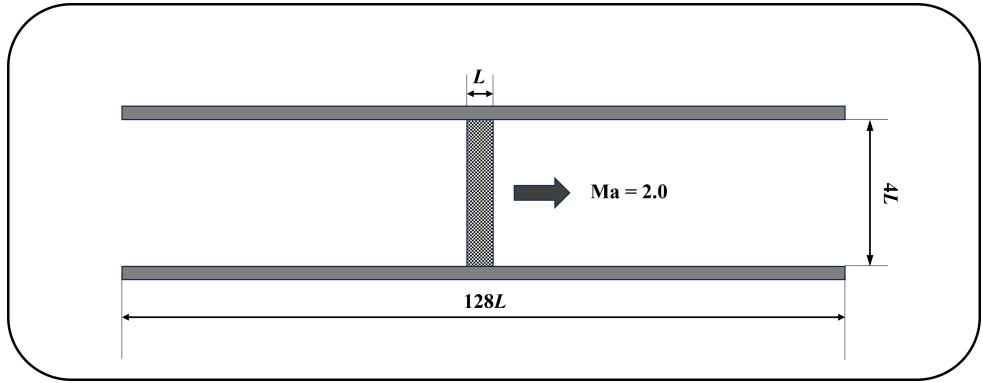

**Fig. 8.** Schematic diagram for the moving piston with Ma 2.0

The flow field ahead of the piston comprises a compression wave, while the flow behind the piston comprises an expansion wave. The problem has an available exact solution (Liepmann and Roshko, 1956). Initially, the hypothetical fluid (molar mass 11640.3) in the entire domain has zero velocity, a pressure of 1, and a temperature of 1. This data provides the speed of sound of unity, a case preferred for validation studies. At the beginning of the simulation, the piston is given a sudden velocity, $u_{piston}$ corresponding to a Mach number of 2.0.

To investigate the grid convergence of the method, the computational domain is discretized using three different uniform grids: 512×16, 1280×40, and 2560×80, corresponding to mesh spacing ($\Delta h$) of $0.25L$, $0.1L$, and $0.05L$, respectively, in both $x$ and $y$ dimensions. The coarsest grid ensures that the piston encompasses at least two finite-volume cells. Simulations are carried out until $t = 25L/u_{piston}$ with RK4 time integration scheme.

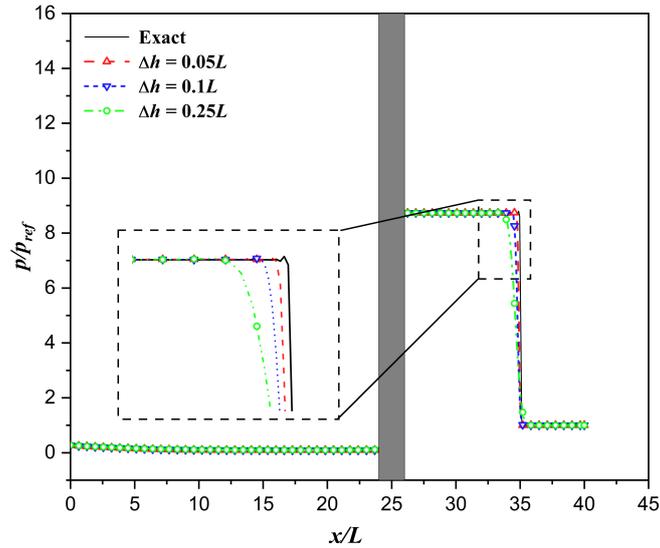

**Fig. 9.** Grid convergence study based on the normalized pressure plotted against normalized distance along centre line

The grid convergence study is shown in Fig. 9, showing the normalized pressure variation across the piston. The shock in front of the piston is visible due to a significant rise in the static pressure. However, due to small pressure changes in the expansion wave, all the plots behind the piston are coincident. The results indicate that the finer grid with a spacing of 0.05$L$ mm yields solutions closer to the exact solution, particularly near sharp gradients. We compared the accuracy of the IBM solver with different flux schemes in Fig. 10 by comparing the normalized pressure and normalized density profiles additionally, the obtained results are compared with exact results. Nearly all flux schemes demonstrate the capability to resolve the shock waves generated by the sudden motion of a body with reasonable accuracy.

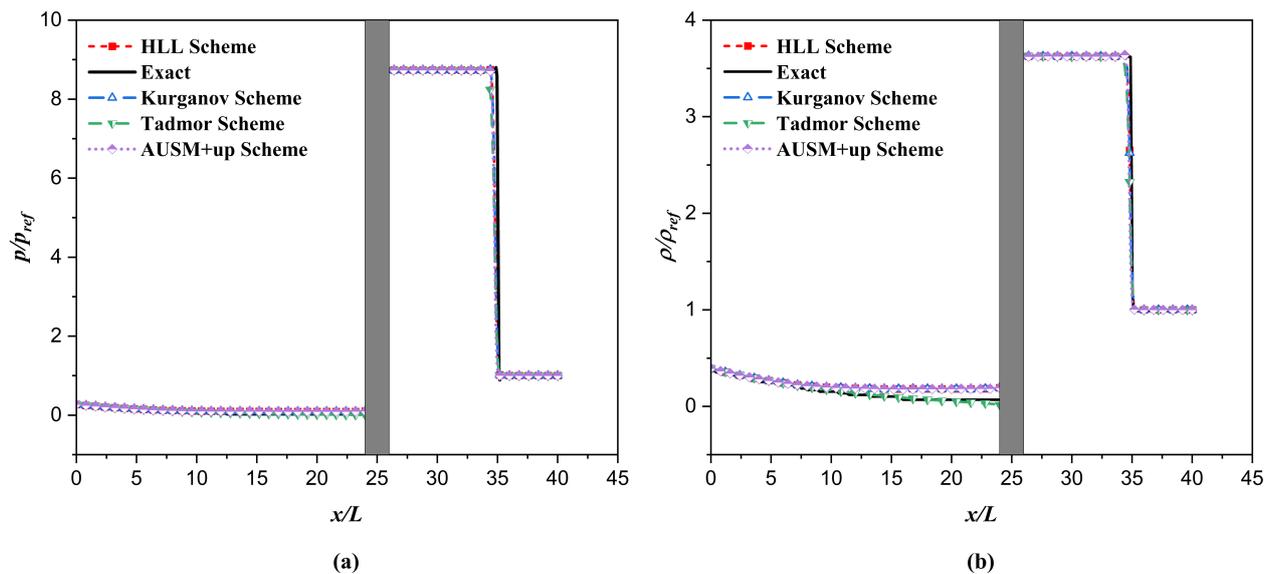

(a)                  (b)

**Fig. 10.** Comparison of (a) normalized pressure (b) normalized density for the moving piston at Ma 2.0 along the centre line using different flux schemes

The moving piston test is also performed for a rotated frame to establish that the developed immersed boundary solver is not sensitive to the orientation of the immersed geometry. We ran the same moving piston problem after rotating the mesh by angle of 45º. The pressure and velocity contours for the rotated frame are shown in Fig. 11(a and b) respectively. It can be seen that the present solver can effectively capture the shock location irrespective of the orientation. Also, the plots of normalized pressure for the two cases are compared in Fig. 12. The location of the shock in the rotated grid case is the same as that of the non-rotated grid, which further confirms that the current method is able to capture the shock independent of the grid rotation.

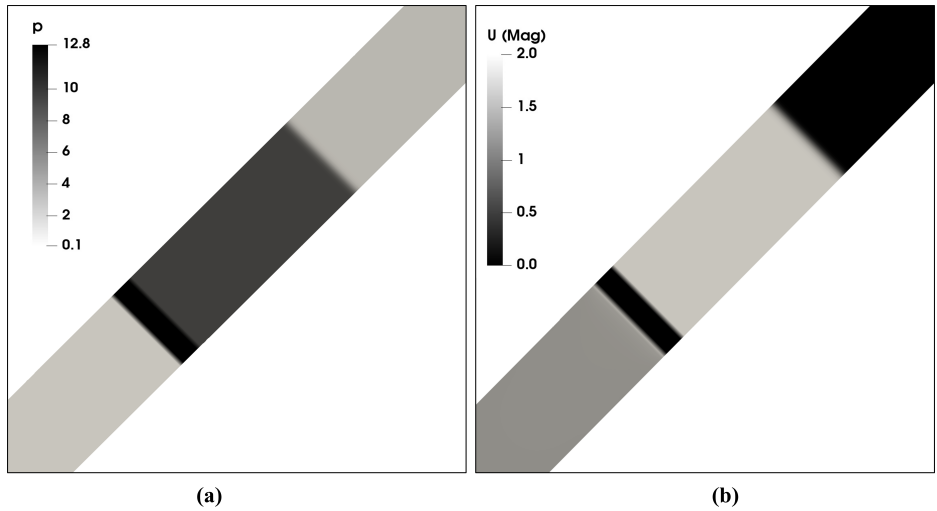

**Fig. 11.** (a) Pressure contour and (b) velocity contour for moving piston at Ma 2.0 in a rotated frame

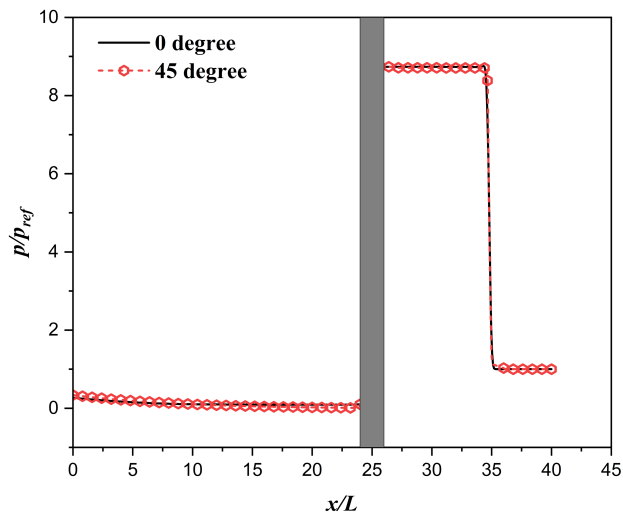

**Fig. 12.** Comparison of pressure contours for a piston moving at Ma 2.0 at two different orientations

### 3.3. Interaction of moving shock with stationary cylinder

This case shows the detailed interaction of moving shock with a stationary cylinder, a well-studied case in literature (Borazjani 2021; Chaudhuri et al. 2010; Yang et al. 1987; Zółtak and Drikakis 1998). The moving shock interacts with the solid cylinder at Ma 2.81. A rectangular domain of size 20$D$

×15$D$ (where $D$ is the diameter of the cylinder) consists of 2100 × 1400 cells in the $x$ and $y$ directions, respectively. The moving shock is initialized at 5$D$. The cylinder is located at 6$D$ from the inlet. The slip boundary condition is implemented at the cylinder together with the top and bottom boundaries of the domain. Figure 13 shows the numerical schlieren obtained over different time steps. The incident shock (IS) is impinging on the cylinder. As a result, a reflected shock (RS) and a Mach shock ($MS_1$) are generated at $t = 0.6$. The point where the RS, $MS_1$, and the IS meet is designated as the first triple point ($TP_1$).

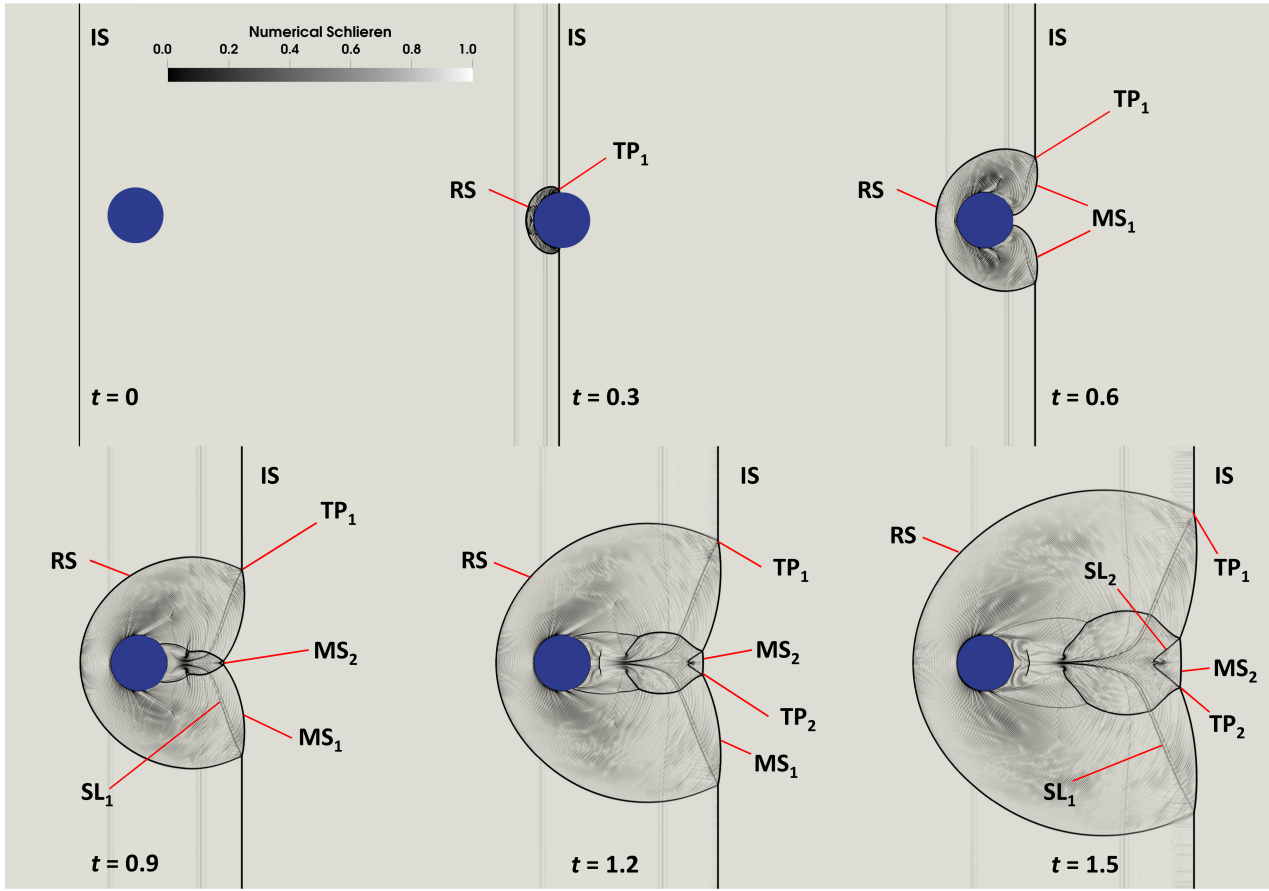

**Fig. 13.** Numerical schlieren of shock diffraction by cylinder

As the Mach shock $MS_1$ propagates downstream, it reflects from the symmetry plane leading to the formation of another Mach wave $MS_2$. Further, the interaction of $MS_2$ with $MS_1$ creates another triple point ($TP_2$), which can be observed after $t = 0.9$. The interaction of these shocks results in the formation of two distinct slip lines ($SL_1$ and $SL_2$). The shocks' interaction and the resulting slip line formation are accurately captured and align well with the schlieren visualization provided by Borazjani (2021). The computed loci of the two triple points are plotted in Fig. 14 for all the flux schemes. It can be noticed that the $TP_1$ originates very close to the surface of the cylinder, whereas the $TP_2$ starts behind the cylinder in the wake region. The comparison seems to be substantially matching for all the schemes and for both the upper and lower systems of shocks. Also, we have compared the results with published data (Borazjani 2021; Chaudhuri et al. 2010). The results seem to follow the published data with a very small deviation far from the cylinder.

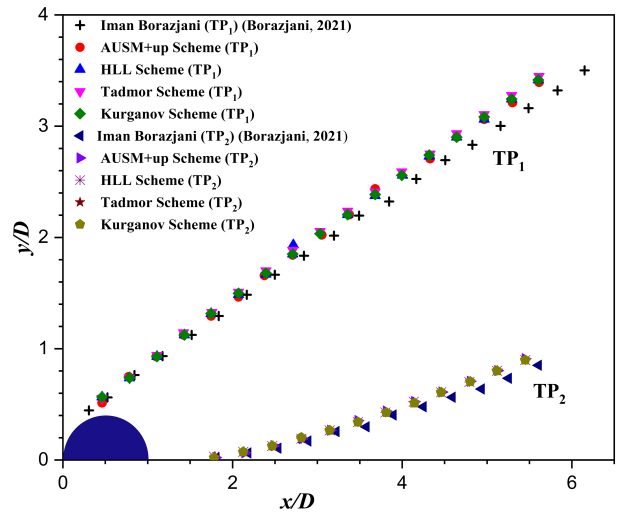

**Fig. 14.** Comparison of the loci of triple points ($TP_1$ and $TP_2$)

### 3.4. Supersonic flow over an aerofoil

In this case, we simulated the inviscid flow over a NACA 0012 aerofoil at Ma 1.5 and zero angle of attack. The computational domain spans a rectangular region with dimensions $(x_1, x_2)$ ϵ $(-5.5c, 6.5c)$ and $(y_1, y_2)$ ϵ $(−6c, 6c)$, where $c$ represents the chord length of the aerofoil. To accurately resolve the sharp gradients near the immersed boundary, the inner region surrounding the aerofoil within $(3c, 2c)$ is refined to include 65,000 cells. This refinement constitutes a significant portion of the total 1,40,000 cells used in the simulation.

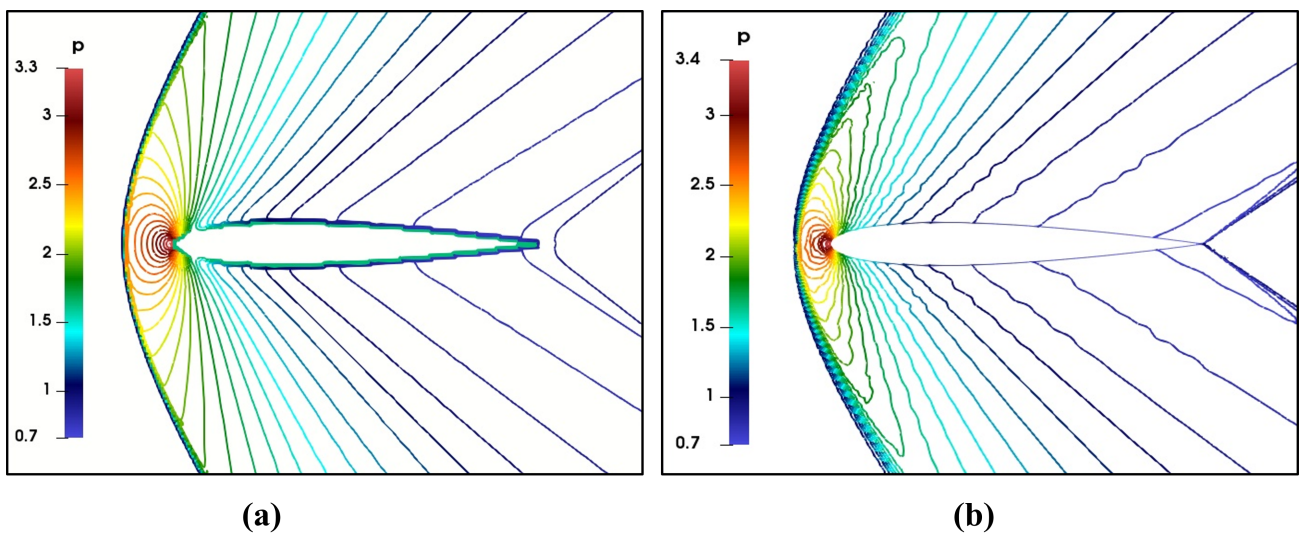

**Fig. 15.** The pressure contour lines (a) current IBM (b) Body-fitted solution

Such a distribution ensures high-resolution coverage in the critical areas close to the aerofoil while maintaining computational efficiency in the less critical outer regions of the domain. The initial conditions are $p=1$ and $T=1$ for the hypothetical gas with the specific gas constant of 0.714, giving the speed of sound of unity. To advance the simulation, an adjustable run time with a Courant number of 0.2 is considered. The pressure contour plot (Fig. 15(a)) shows a sharp accumulation of pressure lines near the aerofoil's leading edge, signifying the presence of a strong shock. There is a good agreement with the body-fitted simulation (Fig. 15(b)), although the body-fitted approach shows slightly more diffused pressure lines near the leading edge. We have compared the non-dimensional coefficient of pressure ($C_p$) over the top surface of the aerofoil using different flux schemes in Fig. 16. All flux schemes closely match the body-fitted results, with the Tadmor and HLL schemes showing the highest agreement. Overall, the immersed boundary solver accurately captures the flow field and shock structure, producing results comparable to conventional body-fitted simulations. Further, the Kurganov scheme shows a significant deviation just before the middle of the chord.

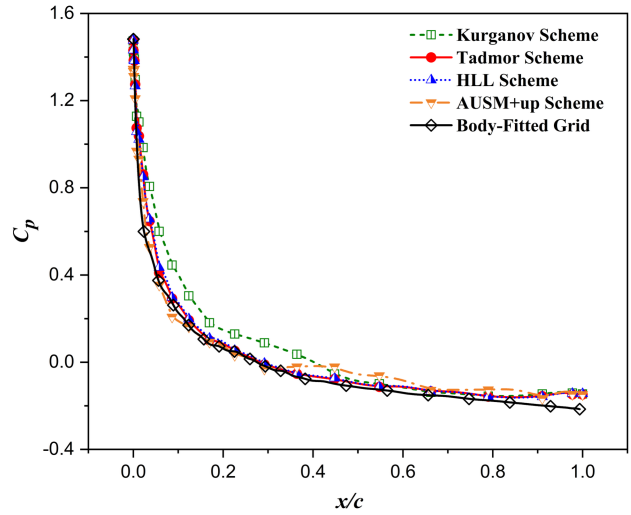

**Fig. 16.** Comparison of surface pressure distribution over the top surface of the aerofoil

### 3.5. Supersonic flow around a 3D Sphere

To evaluate the capability of the IBM solver in accurately predicting the flow field around a three-dimensional bluff body, a sphere with a radius of 23.65 at Ma 5.8 is selected as the test case. The domain is discretized using a three-dimensional Cartesian grid, as depicted in Fig. 17(a), with varying levels of refinement to optimize computational efficiency. After the mesh independence study, we finalized a five-level mesh refinement with a resolution of 0.08 for the innermost layer. Fig. 17(b) compares the density gradient obtained with the simulation with that obtained from the experiment conducted by Machell (1956). The comparison of different flux schemes is shown in the pressure coefficient plots over the sphere's surface in Fig. 18(a). The *x*-axis represents the angular orientation of the surface point, with zero degrees being the stagnation point. All flux schemes align well with experimental data (Machell, 1956), particularly near the stagnation point. Here we have also compared the $C_p$ with the immersed boundary finite volume solver of Brahmachary et al. (2017).

However, all schemes slightly overpredict the pressure coefficient, with HLL being closest to the experimental data. Figure 18(b) compares the shock shape with all schemes, providing results closely matching the Billing correlation (Anderson 2006; Billig 1967). Overall, the HLL and AUSM+up schemes provide robust performance, balancing accuracy and computational efficiency, making them well-suited for high-speed flow simulations around three-dimensional bluff bodies.

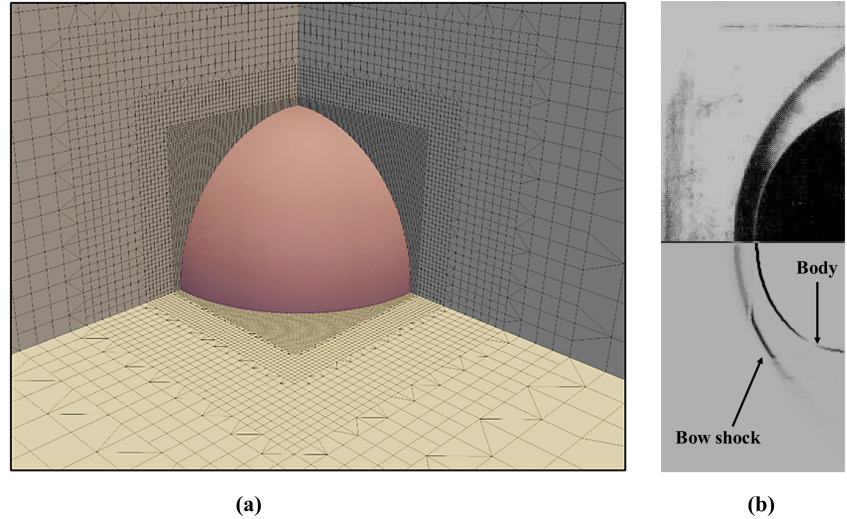

(a)                  (b)

**Fig. 17.** (a) 3-D mesh around cylinder, (b) comparison of shock location obtained for current IBM (bottom) with experimental schlieren (top) for flow over a sphere

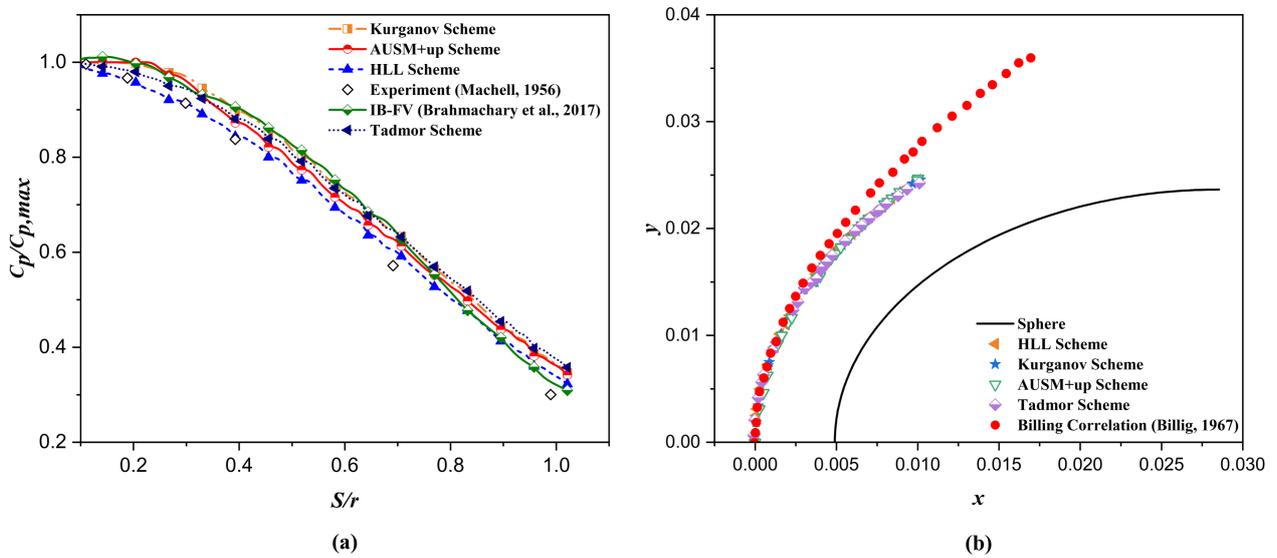

(a)                  (b)

**Fig. 18.** (a) Comparison of normalized surface pressure coefficient for different flux schemes and (b) comparison of shock shape

## 4. Conclusions

This study presents the development of a compressible immersed boundary solver within the OpenFOAM framework, designed for low-supersonic to hypersonic flow regimes. A comprehensive comparison of multiple robust flux schemes was carried out across several benchmark test problems. The solver employs a second-order interpolation technique for imposing boundary conditions, enabling sharp and accurate resolution of complex geometries. In addition, an order of convergence study was performed for smooth supersonic vortex flow, which demonstrates that the solver achieves an overall accuracy close to second-order. A critical outcome of the flux scheme comparison is the observation that no single scheme performs optimally across all conditions, particularly when accounting for varying geometries and Mach number regimes.

The results of the comparative study reveal that the current solver achieves a sharp and accurate resolution of shocks. Furthermore, the numerical solutions exhibit minimal oscillations, a significant achievement given the susceptibility of high-speed flows to numerical oscillations with reconstructed boundary conditions. The solver's capability for non-Cartesian mesh was demonstrated with a test case of a two-dimensional piston moving with supersonic speeds. The extension of the method to three-dimensional simulation of supersonic flow over a sphere further underscores its versatility. Thus, the developed solver not only addresses the challenges of high-speed flow simulations but also provides a flexible, accurate, and robust platform for studying a wide range of flow regimes and geometries.

## Funding resources

The authors acknowledge the financial support provided by DST-SERB through a grant number SER-2339-MID-23-24.

## CRediT authorship contribution statement

Punit Pandey: Data curation, Formal analysis, Investigation, Methodology, Software, Validation, Visualization, Writing – original draft.

Ankit Bansal: Conceptualization, Formal analysis, Investigation, Supervision, Writing – review & editing.

Krishna Mohan Singh: Conceptualization, Formal analysis, Investigation, Supervision, Writing – review & editing.

Yannick Hoarau: Conceptualization, Supervision, Writing – review & editing.

## Declaration of competing interests

The authors declare that they have no known competing financial interests or personal relationships that could have appeared to influence the work reported in this paper.

## Data Availability

The data that supports the findings of this study are available within the article.

# Appendix A. Flux schemes

This appendix provides a brief overview of the different numerical flux schemes used in the solver

## 1. HLL scheme

The HLL (Harten–Lax–van Leer) flux scheme is an approximate Riemann solver for hyperbolic conservation laws to obtain inter-cell fluxes (Toro et al., 1994). It simplifies the wave structure by neglecting the contact discontinuity, focusing only on the fastest left- and right-moving waves for efficient and stable flux computation. The intercell flux $\mathbf{F}_f$ in HLL scheme is given by

$$\mathbf{F}_f = \begin{cases} \mathbf{F}_L & \text{if } 0 \leq S_L \\ \dfrac{S_R \mathbf{F}_L - S_L \mathbf{F}_R + S_L S_R (\mathbf{U}_R - \mathbf{U}_L)}{S_R - S_L} & \text{if } S_L \leq 0 \leq S_R \\ \mathbf{F}_R & \text{if } 0 \geq S_R \end{cases} \quad (A.1)$$

Here, $\mathbf{F}_L$ is the flux calculated from the left state, and $\mathbf{F}_R$ is the flux from the right state. $S_L$ and $S_R$ are the left and right wave propagation speeds given by

$$S_L = \min(u_L - c_L, \tilde{u} - \tilde{c}) \quad (A.2)$$

$$S_R = \max(u_R + c_R, \tilde{u} + \tilde{c}) \quad (A.3)$$

where $u_L$ and $u_R$ are the left and right state face normal velocity components (obtained by taking the dot product of the velocity vector $\mathbf{u}$ with the face normal $\mathbf{n}$), $c_L$ and $c_R$ are the left and right state speed of sound and $\tilde{u}$ and $\tilde{c}$ are the Roe-averaged velocity and speed of sound defined as

$$\tilde{u} = \frac{\left(\sqrt{\rho_L}\, u_L + \sqrt{\rho_R}\, u_R\right)}{\left(\sqrt{\rho_L} + \sqrt{\rho_R}\right)} \quad (A.4)$$

$$\tilde{c} = \left((\gamma - 1)\left(\tilde{H} - \tfrac{1}{2}\tilde{u}^2\right)\right)^{\frac{1}{2}}, \quad (A.5)$$

where $\rho_L$ and $\rho_R$ are the left and right state densities and $\tilde{H}$ is given by

$$\tilde{H} = \frac{\left(\sqrt{\rho_L}\, H_L + \sqrt{\rho_R}\, H_R\right)}{\left(\sqrt{\rho_L} + \sqrt{\rho_R}\right)}, \quad (A.6)$$

in which $H$ represents the enthalpy per unit mass defined as

$$H = \frac{E + p}{\rho} \quad (A.7)$$

## 2. Tadmor/Kurganov scheme

These two different yet similar flux schemes were introduced by Kurganov and Tadmor (2000) and have been implemented in the rhoCentralFoam solver by Greenshields et al. (2009). Both schemes do not require a solution to the Riemann problem, making them computationally efficient. In the Kurganov scheme, the inter-cell flux is formulated as

$$\mathbf{F}_f = (\Omega_L \, \mathbf{U}_L + \Omega_R \, \mathbf{U}_R) + \mathbf{F}_p \tag{A.8}$$

where

$$\Omega_L = u_L \, b_L - b \quad \text{and} \quad \Omega_R = u_R b_R + b \tag{A.9}$$

$$b_L = \frac{b^+}{b^+ + b^-}, \quad b_R = \frac{b^-}{b^+ + b^-}, \quad \text{and} \quad b = \frac{b^+ b^-}{b^+ + b^-} \tag{A.10}$$

where $\mathbf{F}_p$ is the pressure-driven flux across faces

$$\mathbf{F}_p = \begin{bmatrix} 0 \\ (b_L \, p_L + b_R \, p_R) \\ b(p_L - p_R) \end{bmatrix} \tag{A.11}$$

$$b^+ = \max(\max(u_L + c_L, u_R + c_R), 0) \text{ and } b^- = \min(\min(u_L - c_L, u_R - c_R), 0) \tag{A.12}$$

The Tadmor scheme is very similar to the Kurganov scheme; however, it defines

$$b_L = b_R = 0.5 \text{ and } b = \max(|b^-|, |b^+|). \tag{A.13}$$

### 3. AUSM+up scheme

The AUSM+ scheme was introduced by Luo et al. (2003), wherein the fluxes are calculated based on the mass flux across the face. The scheme was further improved by Liou (2006) to include low-speed correction by introducing an additional diffusion term ($M_p$ and $P_u$). The AUSM+up flux is expressed as:

$$\mathbf{F}_f = 0.5[M_* c_* (\mathbf{U}_L + \mathbf{U}_R)] - 0.5[|M_*| c_* (\mathbf{U}_L - \mathbf{U}_R)] + \mathbf{F}_p \tag{A.14}$$

where

$$\mathbf{F}_p = \begin{bmatrix} 0 \\ p_* \\ 0 \end{bmatrix} \tag{A.15}$$

Here, the star variables represent the interface values of the parameters, speed of sound at the cell interface given by

$$c_* = 0.5(c_L + c_R) \tag{A.16}$$

where subscripts $L$ and $R$ represent the contributions from left and right states. The interface Mach number $M_*$ is defined as

$$M_* = \mathcal{M}_{4,L}^+ + \mathcal{M}_{4,R}^- + M_p \tag{A.17}$$

where superscripts $-$ and $+$ correspond to the left and right running waves, respectively. The split Mach numbers $\mathcal{M}_m^\pm$ are the polynomials functions of degree $m = \{1, 2, 4\}$ given by

$$\mathcal{M}_1^{\pm}(M) = 0.5(M \pm |M|) \tag{A.18}$$

$$\mathcal{M}_2^{\pm}(M) = \begin{cases} \mathcal{M}_1^{\pm}(M) & |M| \geq 1 \\ \pm 0.25(M \pm 1) & |M| < 1 \end{cases} \tag{A.19}$$

$$\mathcal{M}_4^{\pm}(M) = \begin{cases} \mathcal{M}_1^{\pm}(M) & |M| \geq 1 \\ \mathcal{M}_2^{\pm}(M)[1 \mp 16\beta \mathcal{M}_2^{\pm}(M)] & |M| < 1 \end{cases} \tag{A.20}$$

Note that $M = \mathbf{u} \cdot \mathbf{n}/c_*$, and has a negative value on the left and a positive value on the right of the interface. The diffusion term $M_p$ is given by

$$M_p = -\frac{K_p}{f_a} \max(1 - \sigma \bar{M}^2, 0) \frac{p_R - p_L}{\rho_* c_*^2} \tag{A.21}$$

where $f_a = 1, \sigma = 1, K_p = 0.25$ are dissipation parameters. The density, $\rho_*$, is computed as the average of the neighboring cell densities $((\rho_L + \rho_R)/2)$. The average Mach number is expressed as

$$\bar{M} = \sqrt{\frac{u_L^2 + u_R^2}{2c_*^2}} \tag{A.22}$$

The interface pressure $p_*$ is given by

$$p_* = \mathcal{P}_{5,L}^+ p_L + \mathcal{P}_{5,R}^- p_R + P_u \tag{A.23}$$

where $\mathcal{P}_{(n)}^{\pm}$ is the pressure splitting polynomial function of degree $n$. The 5th-degree polynomial function yields more accurate solutions and is given by

$$\mathcal{P}_5^{\pm}(M) = \begin{cases} \frac{1}{M}\mathcal{M}_1^{\pm}(M) & |M| \geq 1 \\ \pm \mathcal{M}_2^{\pm}(M)[(2 \mp M) - 16\alpha M \mathcal{M}_2^{\pm}(M)] & |M| < 1 \end{cases} \tag{A.24}$$

with

$$\alpha = \frac{3}{16}(5f_a - 4) \quad \text{and} \quad \beta = \frac{1}{8}. \tag{A.25}$$

and,

$$P_u = -K_u \mathcal{P}_{5,L}^+ \mathcal{P}_{5,R}^- (\rho_L + \rho_L) f_a c_* (u_R - u_L) \tag{A.26}$$

where $K_u = 0.75$. The incorporation of the diffusion term $M_p$ becomes important for regions of low Mach numbers, where numerical checker boarding may affect the accuracy of the solution

### Appendix B. Immersed boundary solution procedure

The solver proceeds by reconstructing the flow variables near the immersed boundary and solving the governing equations to update the flow field at each time step. For moving bodies, the boundary is updated during the simulation, after which the reconstruction and solution steps are repeated; the detailed algorithm is described below.

1. The computational cells are classified as fluid, solid, or immersed-boundary cells, and the masking function is assigned.
2. The flow variables pressure, velocity and temperature are initialized, and the initial immersed body is defined.
3. The simulation time is set to $t = 0$, and the time step $\Delta t$ is computed from the CFL condition.
4. Immersed-boundary points, normals, and interpolation stencils are constructed.
5. Flow variables near the immersed boundary are reconstructed using the weighted least-squares IB method to enforce boundary conditions.
6. Numerical fluxes of mass, momentum, and energy are evaluated at all cell faces.
7. The governing conservation equations are advanced in time to update the flow variables.
8. The condition $t > t_b$ is evaluated, where $t_b$ denotes the prescribed onset time of body motion.
9. If the body is moving, its position is updated, the masking and IB geometric data are recalculated, and the IB reconstruction is repeated.
10. The simulation time is updated as $t = t + \Delta t$.
11. The simulation terminates if $t \geq t_{\text{final}}$, or if the residuals fall below a tolerance.